\documentstyle[aps,prl,epsf,floats]{revtex} 
\draft

\begin{document}
\twocolumn[\hsize\textwidth\columnwidth\hsize\csname
@twocolumnfalse\endcsname 


\preprint{UCLA/00/TEP/22; BNL-HET-00/26 
} 

\title{Neutrinos produced by ultrahigh-energy photons at high red shift} 

\author{Alexander Kusenko$^{1,2}$ and Marieke Postma$^1$
}
\address{$^1$Department of Physics and Astronomy, UCLA, Los Angeles, CA
90095-1547 \\ $^2$RIKEN BNL Research Center, Brookhaven National
Laboratory, Upton, NY 11973 }

\date{July, 2000}

\maketitle
             
\begin{abstract}
  Some of the proposed explanations for the origin of ultrahigh-energy
  cosmic rays invoke new sources of energetic photons ({\em e.g.},
  topological defects, relic particles, {\em etc.}).  At high red shift,
  when the cosmic microwave background has a higher temperature but the
  radio background is low, the ultrahigh-energy photons can generate
  neutrinos through pair-production of muons and pions.  Slowly evolving
  sources produce a detectable diffuse background of $10^{17}$eV neutrinos.
  Rapidly evolving sources of photons can be ruled out based on the
  existing upper limit on the neutrino flux.
    
\end{abstract}

\pacs{PACS numbers: 98.70.Sa, 95.85.Ry, 98.70.Vc  \hspace{1.0cm}
BNL-HET-00/26; UCLA/00/TEP/22} 

\vskip2.0pc]

\renewcommand{\thefootnote}{\arabic{footnote}}
\setcounter{footnote}{0}

Discovery of cosmic rays~\cite{data} with energies beyond the
Greisen-Zatsepin-Kuzmin (GZK) cutoff~\cite{gzk} presents an outstanding
puzzle in astrophysics and cosmology~\cite{reviews}.  Many proposed
explanations invoke a new source, such as superheavy relic
particles~\cite{particles,gk2,kt} or topological
defects~\cite{TD_review,TD,TD_ber,TD_nu}, that can generate photons at both
low and high red shifts.  In understanding the origin of the
ultrahigh-energy cosmic rays (UHECR), it is crucial to distinguish such
sources from more conventional astrophysical ones~\cite{conv,mpr}.  In this
{\em letter} we show that a diffuse background of neutrinos with energies
$\sim 10^{17}$eV can be generated by ultrahigh-energy photons at high red
shift.

Generation of ultrahigh-energy neutrinos has been
studied~\cite{TD,TD_ber,TD_nu,conv,mpr,hs,stecker,ysl,pj,ps,whitepaper}
for various sources at small red shift, for which muon pair-production can
be neglected.  However, a substantial flux of neutrinos could be produced
at earlier times, when the propagation of photons was different from that
in the present universe because the intergalactic magnetic field was
weaker, the density of radio background was lower, and the cosmic microwave
background (CMB) density and temperature were higher.

At red shift $z$ the cosmic microwave background radiation (CMBR) has
temperature $T_{_{CMB}}(z) = 2.7 (1+z){\rm K}$.  Because of this, at
high red shift photon-photon and electron-photon interactions can
produce pairs of muons and charged pions, whose decays generate
neutrinos. This is in sharp contrast with the $z
\stackrel{<}{_{\scriptstyle \sim}} 1$ case, where the photons do not
produce neutrinos as they lose energy mainly by scattering off the
radio background through electron-positron pair production and
subsequent electromagnetic cascade~\cite{berezinsky,reviews}.  The
ratio of the CMBR density to that of universal radio background (RB)
increases at higher $z$, and the processes $\gamma
\gamma_{_{CMB}}\rightarrow \mu^+ \mu^-$ and $e
\gamma_{_{CMB}}\rightarrow e \mu^+ \mu^- $ can produce muons, which
decay into neutrinos: $\mu \rightarrow e \nu_e \nu_{\mu}$.  The
threshold for these interactions is $\sqrt{s}> 2 m_\mu=0.21$GeV, or
\begin{equation}
E_{\gamma,e} > E_{\rm th}(z)=\frac{ 10^{20}{\rm eV}}{1+z}
\label{threshold}
\end{equation}

Our discussion applies to any source of photons active at high red shift.
The latter requirement excludes some astrophysical sources~\cite{conv}.
Topological defects~\cite{TD_review,TD,TD_nu} and decaying relic
particles~\cite{particles,gk2}, however, could operate even at $z\gg1$.
These sources are expected to produce photons with energies as high as
$10^{20}$eV.  We will describe a neutrino signature of this class of
sources.

At $z<1$ the main source of energy loss for photons is electromagnetic
cascade that involves $e^+e^-$ pair production (PP) on the radio
background photons.  The radio background is generated by normal and
radio galaxies.  Its present density~\cite{biermann} is higher than
that of CMB photons in the same energy range.  The radio background
determines the mean interaction length for the $e^+e^-$ pair
production.  At red shift $z$, however, the density of CMB photons is
higher by a factor $(1+z)^3$, while the density of radio background is
either constant or, more likely, lower.  Some models of cosmological
evolution of radio sources~\cite{condon} predict a sharp drop in the
density of radio background at red shift $z
\stackrel{>}{_{\scriptstyle \sim}} 2$.  More recent
observations~\cite{recent} indicate that the decrease of radio
background at $z>2$ is slow.  However, one expects the CMB to become a
more important source of energy losses for photons at higher $z$
because of the $(1+z)^3$ increase in the density of CMB photons.  Let
$z_{_R}$ be the value of red shift at which the scattering of
high-energy photons off CMBR dominates over their scattering off RB.
Based on the analyses of Refs.~\cite{condon,recent}, we take $z_{_R}
\sim 5$.  Another source of energy losses in the electromagnetic
cascade is the synchrotron radiation by the electrons in the
intergalactic magnetic field (IGMF).  This is an important effect for
red shift $z<z_{_M}$, where $z_{_M} \sim 5$ corresponds to the time
when the synchrotron losses are not as significant as the interactions
with the CMB radiation.  We will use the value $z_{\rm min}=\max
(z_{_R},z_{_M}) \approx 5$ in what follows.  As discussed below, a
higher value of $z_{\rm min}$, even as high as 10, would not make a
big difference in the flux of the signature neutrinos.

Let us now consider the propagation of UHE photons at $z>z_{\rm min}$. In
particular, we are interested in neutrino-generating processes, that is,
reactions that produce muons and pions.  For photon energies above the
threshold for muon pair production~(\ref{threshold}), the reactions $\gamma
\gamma_{_{CMB}}\rightarrow e^+ e^-$, $\; \gamma \gamma_{_{CMB}}\rightarrow
e^+ e^- e^+ e^-$ and $\gamma \gamma_{_{CMB}}\rightarrow\mu^+ \mu^-$ are
possible.  For $\sqrt{s}> 2m_{\pi^{\pm}} = 0.28$GeV the charged pion
production may also occur.  Among the processes listed above, the electron
pair production (PP) has the highest cross section for photon energies
$E_\gamma \stackrel{<}{_{\scriptstyle \sim}} 5 \times 10^{20} {\rm eV} 
/(1+z)$. Since the energies of the two interacting photons are vastly
different, either the electron or the positron from PP has energy close to
that of the initial photon.  At higher photon energies, double pair
production (DPP) becomes more important~\cite{dpp}. Four electrons, each
carrying about $1/4$ of the initial photon energy, are produced in this
reaction.  Thus, after an initial $\gamma \gamma_{_{CMB}}$ interaction one
ends up with one or more UHE electrons.

These electrons continue to scatter off CMBR. At lower energies, inverse
Compton scattering (ICS), $e \gamma_{_{CMB}}\rightarrow e \gamma $,
converts high-energy electrons into high-energy photons~\cite{reviews}.
However, at energies above the muon threshold, higher order processes, such
as triplet production (TPP) $e \gamma_{_{CMB}}\rightarrow e\, e^+ e^-$ and
muon electron-pair production (MPP) $e \gamma_{_{CMB}}\rightarrow e\, \mu^+
\mu^-$, dominate.  For center of mass energies $s \gg m_e^2$, the
inelasticity $\eta $ for TPP is very small: $\eta \simeq 1.768
(s/m_e^2)^{-3/4} < 10^{-3}$~\cite{tpp,tpp2}.  One of the electrons produced
through TPP, carries almost all ($1-\eta$) of the incoming electron's
energy.  It can interact once again with the CMBR. As a result, the leading
electron can scatter many times before losing a considerable amount of
energy. Hence, the energy attenuation length $\lambda_{\rm eff}$ is much
greater than the TPP interaction length: $\lambda_{_{\rm TPP}} \simeq \eta
\lambda_{\rm eff}$.

To see if neutrinos are produced, one must compare this {\em energy
attenuation} length with the {\em interaction length } for muon pair
production in processes like 
$e \gamma_{_{CMB}}\rightarrow e\, \mu^+ \mu^-$.  Above the pion threshold,
pion production is yet another channel that drains the energy out of the
electromagnetic cascade and into neutrinos.   We note that
even a single neutrino-producing channel is enough for UHE photons to
produce neutrinos at high red shift.  The fact that there are
several such channels makes little difference.  

Let us compare the TPP energy attenuation length $\lambda_{\rm eff}$ with
the interaction length for muon pair production.  The interaction length is
given by $ \lambda^{-1} \simeq \left< n_{_{CMB}} \right> v \sigma$, and
thus the ratio is $R=\lambda_{\rm eff}/\lambda_{_{\rm MPP}} \simeq
\sigma_{_{\rm MPP}}/(\eta \sigma_{_{\rm TPP}})$. For $s \gg m_e^2$ the
cross section for TPP is~\cite{tpp,tpp2}
\begin{equation} 
\sigma_{_{TPP}} \simeq \frac{3\alpha}{8
\pi}\sigma_{_{T}} \left( \frac{28}{9} {\rm ln}\frac{s}{m_e^2}
-\frac{218}{27} \right), 
\end{equation} 
where $\sigma_{_{\rm T}}$ is the Thompson cross section.  The MPP cross
section in the energy range just above the threshold $5 m_\mu^2<s<20
m_\mu^2$ is of the order of $0.1-1{\rm mb}$, and the ratio $R \sim 100$.

Since $\lambda_{\rm eff} \gg \lambda_{_{\rm MPP}}$, in the absence of dense
radio background and intergalactic magnetic fields, all electrons with $E>
E_{\rm th}$ pair-produce muons before their energy is reduced by the
cascade. For muon production close to the threshold, each muon carries on
average $1/4$ of the incoming electron's energy~\cite{tpp2}. Muons decay
before they can interact with the photon background.  Each energetic muon
produces two neutrinos and an electron.  The electron produced alongside
the muon pair gets half or more of the incoming electron's energy; it can
interact again with the CMBR to produce muons. This process can repeat
until the energy of the regenerated electron decreases below the threshold
for muon pair production.  Higher energy electrons with energies $E>
2\times 10^{20} {\rm eV}/(1+z)$eV can also produce pions through the
reaction $e \gamma_{_{CMB}}\rightarrow e \pi \pi$. Charged pions decay into
neutrinos, while neutral pions reproduce photons.  As explained above, it
makes little difference through which channel the neutrinos are produced --
as long as there is at least one reaction with a shorter mean free path
than the energy attenuation length.

One can parameterize the rate of photon production as
$\dot{n}_{_X} =\dot{n}_{_\gamma,0} (t/t_0)^{-m}$, with $m=0$ for
decaying relic particles, $m=3$ for ordinary string and necklaces, and
$m \ge 4$ for superconducting strings~\cite{TD_review,TD,TD_ber}.  Let
$z_{\rm max}$ be the red shift at which the universe becomes opaque to
ultrahigh-energy neutrinos. Its value is determined by the neutrino
interactions with the relic neutrino background.  The absorption red
shift for neutrinos with energy $\sim 10^{17}$eV is $z_{\rm max}\sim 3
\times 10^3$~\cite{ggs}. All neutrinos coming from red shift $z_{\rm
min}<z<z_{\rm max}$ contribute to the present flux.  The neutrino flux
is
\begin{eqnarray} 
n_\nu & = & \xi \int^{z_{\rm max}}_{z_{\rm min}} dt \ \dot{n}_{\gamma}(z) \
(1+z)^{-4} \nonumber \\ 
& = & \xi \frac{3}{-2a} \dot{n}_{\gamma,0} \, t_0
\, [(1+z_{\rm min})^a-(1+z_{\rm max})^a] ,
\label{nu_z} 
\end{eqnarray} 
where $\xi $ is the number of neutrinos produced per UHE photon, and
$a=(3m-11)/2$.  We take $\xi \approx 4$ because one photon produces one UHE
electron, which generates a pair of muons, which decay into four neutrinos.
This is probably an underestimate because the remaining electron may have
enough energy for a second round of muon pair-production.  Also, pion
decays produce three neutrinos each. 

For $m<11/3$, $a< 0$, and, according to~eq.~(\ref{nu_z}), most of
neutrinos come from red shift $z\sim z_{\rm min}\approx 5$. All these
neutrinos are produced by photons with energies $E_\gamma > E_{\rm
min} = 10^{20}{\rm eV}/(1+z_{\rm min}) \sim 2 \times 10^{19}$eV.  If
decaying TD's or relic particles are the origin of the UHECR today,
one can use the observed UHECR flux to fix the overall normalization
constant $\dot{n}_{\gamma,0}(E>E_{\rm min})$.  We will use the photon
fluxes calculated in~\cite{TD_ber,pj,ps}.  For various sources, $
\dot{n}_{\gamma,0}(E>E_{\rm min})\sim L^{-1} \int_{E_{\rm min}} 
{\rm d}E \, J(E) $, 
where is $J(E)$ the differential photon flux, and $L$ the length scale
from which the photons are collected.  Because the photon flux is a sharply
falling function of energy, $\dot{n}_{\gamma,0}$ is dominated by
photons with energies $E\sim E_{\rm min}$. Relic particles $(m=0)$ and
monopolonia $(m=3)$ cluster in galaxies and have $L \sim  L_{\rm gal} \sim
100{\rm kpc}$, the size of our galaxy.  Their over-density in the galaxy is
$\sim 2 \times 10^{5}$~\cite{TD_ber}.  Necklaces $(m=3)$ on the other hand are
distributed uniformly throughout the universe; for them $L = L_\gamma
\sim 5$Mpc, the photon absorption length at these energies.

Using the photon flux from Refs.~\cite{TD_ber,pj,ps} one obtains 
the neutrino flux:
\begin{equation}
\phi_{\nu} \sim \left \{
\begin{array}{lll}
 10^{-21}{\rm cm}^{-2} {\rm s}^{-1} {\rm
  sr}^{-1}, & \quad {\rm relic \: particles}\, (m=0), \\ 
 10^{-18}{\rm cm}^{-2} {\rm
  s}^{-1} {\rm sr}^{-1},  & \quad {\rm monopolonium} \, (m=3) , \\  
10^{-16}{\rm cm}^{-2} {\rm s}^{-1} {\rm
  sr}^{-1}, &\quad {\rm necklaces}\, (m=3). \\
\end{array} \right. 
\label{flux}
\end{equation}
Taking $z_{\rm min}=10$ reduces the flux of signature neutrinos from
$m=3$ sources only by a factor $2$.

The energy of these neutrinos at red shift $z$ is $E_\nu (z) \sim E_\mu/3$.
It is then further red shifted by a factor $(1+z)^{-1}$.  Assuming a
falling photon spectrum, we expect most of neutrinos to come from photons
near the threshold,~eq.~(\ref{threshold}). We estimate the energy of these
neutrinos after the red shift $E_\nu \sim 10^{17}$eV. 

If the source in question is a slowly decaying relic particle or some other
source with $m=0$, the neutrinos produced at high red shift are probably
not detectable.  Of course, if the relic particles~\cite{gk2} or
topological defects~\cite{TD_nu} produce UHECR through
$Z$-bursts~\cite{Zburst}, neutrinos from $Z$ decays can be detected. 

Can other sources produce a comparable flux of neutrinos at
$10^{17}-10^{18}$eV?  The neutrino flux $ \phi_{\nu} \sim 10^{-16} {\rm
cm}^{-2} {\rm s}^{-1}{\rm sr}^{-1}$ at $E_\nu \sim 10^{17}$eV exceeds the
background flux from the atmosphere and from pion photoproduction on CMBR
at this energy~\cite{hs,stecker,sp}, as well as the fluxes predicted by a
number of models~\cite{whitepaper}. TD can produce a large flux of primary
neutrinos.  However, the primary flux peaks at $E_\nu\sim 10^{20}$eV, while
the secondary flux peaks at $E_\nu\sim 10^{17}$eV and creates a distinctive
``bump'' in the spectrum.  Models of active galactic nuclei (AGN) have
predicted a similar flux of neutrinos at these energies~\cite{mpr}.  The
predictions of these models have been a subject of debate~\cite{wb}.
However, every one agrees that AGN cannot produce neutrinos with energies
of $10^{20}$eV~\cite{rm}.  So, an observation of $10^{17}$eV neutrinos
accompanied by a comparable flux of $10^{20}$eV neutrinos would be a
signature of a TD rather than an AGN.

There is yet another interesting possibility.  TD with $m \ge 4$, {\em
e.g.}, superconducting strings, cannot give a large enough flux of UHECR
because of the EGRET bound on the flux of $\gamma$-photons~\cite{reviews}.
However, this does not mean they did not exist in the early
universe. Neutrinos with energy $10^{17}$eV are probably the  only 
observable signature of some rapidly evolving sources that could be 
active at high red shift but would have ``burned out'' by now.

To summarize, we have shown that sources of ultrahigh-energy photons that
operate at red shift $z \stackrel{>}{_{\scriptstyle \sim}}5$ produce
neutrinos with energy $E_\nu \sim 10^{17}$eV.  The flux depends on the
evolution index $m$ of the source.  A distinctive characteristic of this
type of neutrino background is a cutoff below $10^{17}$eV due to the
universal radio background at $z<z_{\rm min}$. Detection of these neutrinos
can help understand the origin of ultrahigh-energy cosmic rays.

We thank J. Alvarez-Muniz, P.~Biermann, V.~Berezinsky, F.~Halzen, and
G.~Sigl for helpful comments.  This work was supported in part by the US
Department of Energy grant DE-FG03-91ER40662, Task C, as well as by an
Assistant Professor Initiative grant from UCLA Council on Research.
A.K. thanks CERN Theory Division for hospitality during his stay at CERN
when part of this work was performed.


\end{document}